# Toward a Unified Security Framework for AI Agents: Trust, Risk, and Liability

Jiayun Mo, Xin Kang, *Senior Member, IEEE,* Tieyan Li, *Member, IEEE,* Zhongding Lei, *Senior Member, IEEE*

*Abstract*— **The excitement brought by the development of AI agents came alongside arising problems. These concerns centered around users' trust issues towards AIs, the risks involved, and the difficulty of attributing responsibilities and liabilities. Current solutions only attempt to target each problem separately without acknowledging their inter-influential nature. The Trust, Risk and Liability (TRL) framework proposed in this paper, however, ties together the interdependent relationships of trust, risk, and liability to provide a systematic method of building and enhancing trust, analyzing and mitigating risks, and allocating and attributing liabilities. It can be applied to analyze any application scenarios of AI agents and suggest appropriate measures fitting to the context. The implications of the TRL framework lie in its potential societal impacts, economic impacts, ethical impacts, and more. It is expected to bring remarkable values to addressing potential challenges and promoting trustworthy, risk-free, and responsible usage of AI in 6G networks.**

## I. INTRODUCTION

ARTIFICIAL Intelligence (AI) has rapidly evolved from a theoretical concept into an innovative and transformative force. As destructive technologies, AI embodies innovations that "destroy the usage value of established techniques" and generate profound shifts in both industrial and societal landscapes [1]. These technologies, after being embedded in critical systems, have the potential to transform entire sectors and workflows, driving substantial changes in how businesses, governments, and individuals operate.

However, problems related to AI arise simultaneously alongside their immense development, cautioning the public and delaying the progress of transitioning into an AI-dependent future. One such example would be trust issues centered around AI. Despite the increased presence of AI agents as voice assistants in mobile phones, like Siri, they have a low utilization rate due to users' lack of trust towards their competency and integrity [2]. In addition, moral and ethical concerns are also prominent. Deciding which stakeholders shall be held accountable further complicates the problem. Researchers use the most commonly discussed autonomous vehicles as examples to conduct hypothetical scenario analysis, debating which stakeholders shall hold which portion of legal liability when accidents occur [3]. These rising problems demand solutions, in order to put the advancing technologies into effective uses.

Despite the gradual emergence of solutions in recent years, most of them specialized in finding approaches of fostering and enhancing trust for AI agents, methods of addressing safety concerns within AI agents, or metrics to evaluate AI agents, but none of them are working towards creating an exhaustive system of systems that includes all three aspects, as can be seen in the comparative table in Fig. 1. Consequently, the Trust, Risk, Liability (TRL) framework introduced by this paper aims to fill in the gap of existing research, by tying together the interdependent relationships of trust, risk, and liability to propose a framework for building and enhancing trust, analyzing and mitigating risks, and allocating and attributing liabilities.

## II. INTRODUCTION TO AI AGENTS

The increasing development of AI agent has often generated confusion, even among experts, who may refer to AI agent to talk about very different things [4]. In the mainstream definition underpinned by Russell and Norvig, an agent is defined as anything, humans or robots, that is capable of "perceiving its environment through sensors and acting upon that environment through actuators" [5]. Intelligent agents can be classified into four main types. Firstly, a simple reflex agent makes decisions only based on the present state. A model-based reflex agent is guided by both the current percept and the percept histories it kept track of. A goal-based reflex agent selects actions in order to best reach its goals. Lastly, a utility-based reflex agent aims to maximize its utility according to the utility function. AI agents can be a mixture of the four types of intelligent agents, but most importantly, AI agents are all learning agents, constantly absorbing new information to make better decisions and improve their capabilities [5]. The AI agents discussed in this article are learning agents with LLMs laying the foundation, followed by the operating systems, and finally, the specific AI agents as the top layer. The AI agent contains a profiling module, a memory module, a planning module, and an action module [6]. The profiling module identifies and understands its role and identity [6]. The memory module memorizes all the past percepts and uses them as a basis for experience gathering and iterative learning [6]. The planning module breaks complex tasks into more conquerable subtasks and enables planning processes both with and without external feedback [6]. The action module turns decisions made by the AI agent into specific output results and interacts with the environment accordingly [6].

| Study | Aspects Covered | Integration Level | Applicable Sectors | Novelty of the LRT Framework |
|---|---|---|---|---|
| Chan et al. (2024) | Risk, Liability | High | Public sector, software ecosystems. | Whereas the other studies only touched on at most two of the three aspects, sometimes with partial integration, the LRT framework is fully integrating the three aspects of trust, risk, and liability. The framework also has a broad applicability, ensuring that it may be applied to any domain. |
| Vetrò et al. (2019) | Trust, Risk | Partial | Cross-industry | |
| Russell and Norvig (2010) | Trust, Liability | Partial | Broad applicability | |
| This Study Mo et al. (2024) | Trust, Risk, Liability | High | Broad applicability | |

Figure 1. Comparison of state-of-the-art works.

AI agents, characterized as systems capable of pursuing complex goals with minimal supervision, have far-reaching implications across industrial and societal domains. They align with intentional systems, as conceptualized by Dennett, where their behavior can be interpreted in terms of beliefs and desires [7]. The increasing reliance on these agents necessitates addressing issues of visibility, accountability, and governance. Visibility, defined as "information about where, why, how, and by whom certain AI agents are used," is particularly critical [8]. Overall, these fundamentals about the types and modules of AI agents, as well as how it interplays on a macro scale, are crucial for understanding and using the TRL framework in application scenarios related to AI agents.

## III. INTRODUCTION TO THE TRUST, RISK, LIABILITY FRAMEWORK

The proposed TRL framework ties a liability system, a risk system, and a trust system closely together into an inter-influential cyclic structure. The three systems are each showcased as a wheel in Fig. 2, providing an overarching guidance for the development and utilization of responsible, risk-free, and trustworthy AI agents. Each system has an interior circle whose shape resembles the Chinese philosophy concept of Yin Yang, referring to two opposite yet interconnected forces that co-exist and contribute simultaneously to the overall balance of the system. Here, each wheel contains two parts that also contribute their individual, interdependent shares in balancing the system. Whereas the inner circle outlines a general framework, the exterior circle provides extensions that may be specifically applied to different domains in which the systems are to be used, which is AI in this context.

The trust system on the top has an interior circle that is separated into societal trust and technical trust, which respectively focuses on societal factors and technical functionalities that foster stronger trustworthiness. Whereas trust may be about forming a subjective relationship between the trustor and the trustee, trustworthiness is an objective measure of the extent that one can be trusted [9]. The exterior circle highlights the three core requirements for a trustworthy AI Agent, namely dependable, controllable, and alignable. The liability system is composed of attribution models that correspond responsibilities with stakeholders and accountability mechanisms that allocate liabilities to stakeholders. The European Union (EU) AI Act and the European Union Liability Rules for AI in the exterior circle are two such extensions useful for responsibility attribution and liability allocation of AI agents. The bottom right risk system features the risk analysis and risk management segments. The former conducts systematic analysis of potential risks, while the latter proposes strategies to cope with the risks. Extensions that are useful in the context of AI agents, listed in the exterior circle, include the Germany AI Risk Levels, EU AI Act

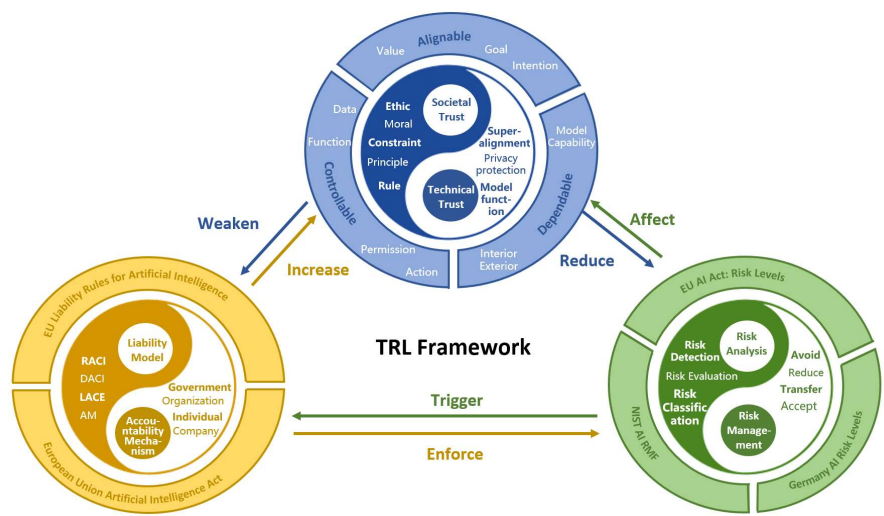

Figure 2. The Trust, Risk, Liability framework.

Risk Levels, and the National Institute of Standards and Technology (NIST) AI Risk Management Framework (RMF).

The three systems are interdependent and interconnected, forming the TRL framework. Each system affects and is affected by the others; their relationship is characterized by the keyword labelled on the arrows in Fig. 2, which will be explained below.

Firstly, responsibility affects trust. Psychology studies showed a positive correlation between responsibility and trust, suggesting that more responsible agents are generally trusted more [10]. Meanwhile, trust reduces the dependency on accountability mechanisms. Higher trustworthiness implies that there can be less precautious measures, and thus less need for such mechanisms. Secondly, risk triggers accountability mechanisms. Whilst risks always exist, the accountability mechanisms are only triggered after the actual occurrence of the event. Higher risks need to correspond to higher liability, so that people will be motivated to conduct morally, legally, and ethically right actions [11]. Meanwhile, the accountability mechanisms help to enforce the risk management process. Stakeholders with more responsibilities will strive to manage risks with higher efforts, in order to ensure that they do not need to bear the consequences. Clear attribution models and effective accountability mechanisms can facilitate risk management [12]. Lastly, risk level affects the degrees of trust. Agents generally make trust decisions less for high-risk events, due to the immense loss that will be incurred by a wrongfully placed trust; meanwhile, trust decisions regarding low-risk events are made with more ease [13]. At the same time, trust reduces the cost of risk management. With higher trust from the agents, less measures to control and contain the risks are required.

Overall, the three interdependent systems work together to form the TRL framework. The next section will respectively introduce the trust, risk and liability system in more details.

## IV. TRUST SYSTEM

The trust system within the TRL framework consists of four establishment dimensions, three core trust requirements, two trust aspects, and one purpose, as shown in Fig. 3.

The four establishment dimensions of trust include the trustee's name, exterior, behavior, and interior. The name's trustworthiness is influenced by its cultural significance, as well as the ease to be memorized, pronounced, and recognized. Appearance, sound, expression, and style are exterior factors determining trust levels. Next, the perceived accountability and traceability of a trustee is crucial for maintaining trust, which points to the importance of the behavior dimension [14]. Behaviors including the trustee's reaction, response, communication, decision making processes, functional optionality, and data processing abilities all influence trust. Lastly, value, characteristic, logic, knowledge, capability, and goal are all crucial interior factors for enhancing trust.

The aforementioned dimensions should be designed while keeping the three core trust requirements in mind, which are dependability, controllability, and alignability. Model capability, along with the name and exterior dimensions, determine trustee's dependability; data, function, permission, and the behavior dimension shape controllability; and value, goal, and intention are related to alignability.

Moving on to the two trust aspects, factors of societal trust include culture, belief, moral, transparency, risk, familiarity, brand, value, worldview, and ability, while technical trust includes identity, data, algorithm, copyright, privacy, fairness, hardware, explainability, anti-counterfeiting, and rejection.

Societal trust centers around social factors that play a role in influencing trust levels, given the differing social environment trustors are located in and the social interactions they grew up with [15]. O'Neill has confirmed the idea that external governance structures and societal context play a role in shaping trust. On the other hand, factors from the technical aspect also contribute to a trustee's perceived trust levels, since trust is often based on the perceived reliability and competence of the system.

Overall, factors from the four establishment dimensions should be designed in a way that strives to satisfy the three core requirements; the two trust aspects are used to foster trust, reaching the ultimate purpose of consistency between words and deeds.

## V. RISK SYSTEM

Risk exists in a system regardless of whether an event has taken place and whether a scenario has already occurred. Identifying, pinpointing, and mitigating risks are thus of crucial importance to the smooth functioning of a system. The process is mainly broken down into risk analysis and risk management with two proposed approaches; one approach utilizes ranking models, while the other one relies more heavily on mathematics and probability models, as shown in Fig. 3.

### A. RANKING-BASED RISK ANALYSIS

The ranking-based risk analysis procedure starts with choosing a risk ranking model and setting suitable trust indicators. Ranks and risk indicators may differ based on different scenarios and the overall scale of risks involved. The next step is to evaluate risks associated with each scenario against different indicators. A table can be used to better facilitate the evaluation process; the rows are for the scenarios while the columns are for the risk indicators, and each cell documents whether the indicator is satisfied. Lastly, a risk mitigation strategy should be determined based on the risk ranking result. Risk avoidance should be adopted for scenarios or events corresponding to the highest level of risks. Risk transfer is suitable for high-risk events. Low risk scenarios need risk reduction strategies. Risk acceptance is for events corresponding to the lowest level of risks. Methods related to each strategy are listed in the Risk System in Fig. 3. Overall, an ideal ranking-based model should be easily definable, extensible, and standardizable.

### B. MATH-BASED RISK ANALYSIS

Besides ranking-based risk analysis models, another approach utilizes math-based models to analyze and manage risks. The models should aim to be modellable, quantifiable, and computable. The procedure is similar to the ranking-based approach introduced above. The first step is choosing a risk analysis model and building the model. Examples of suitable models include Monte-Carlo simulation, Bayesian network, and event tree analysis. After mathematical and statistical evaluations, the risk analysis model generates insights for the potential risks involved in the scenario. Based on the risk level, one of the four aforementioned risk mitigation strategies, namely risk avoidance, risk transfer, risk reduction, and risk acceptance, will be chosen to manage the risks. Overall, with the help of ranking-based and math-based risk analysis models and risk mitigation strategies, the risk system aims to ensure that risks are recognizable, evaluable, and controllable.

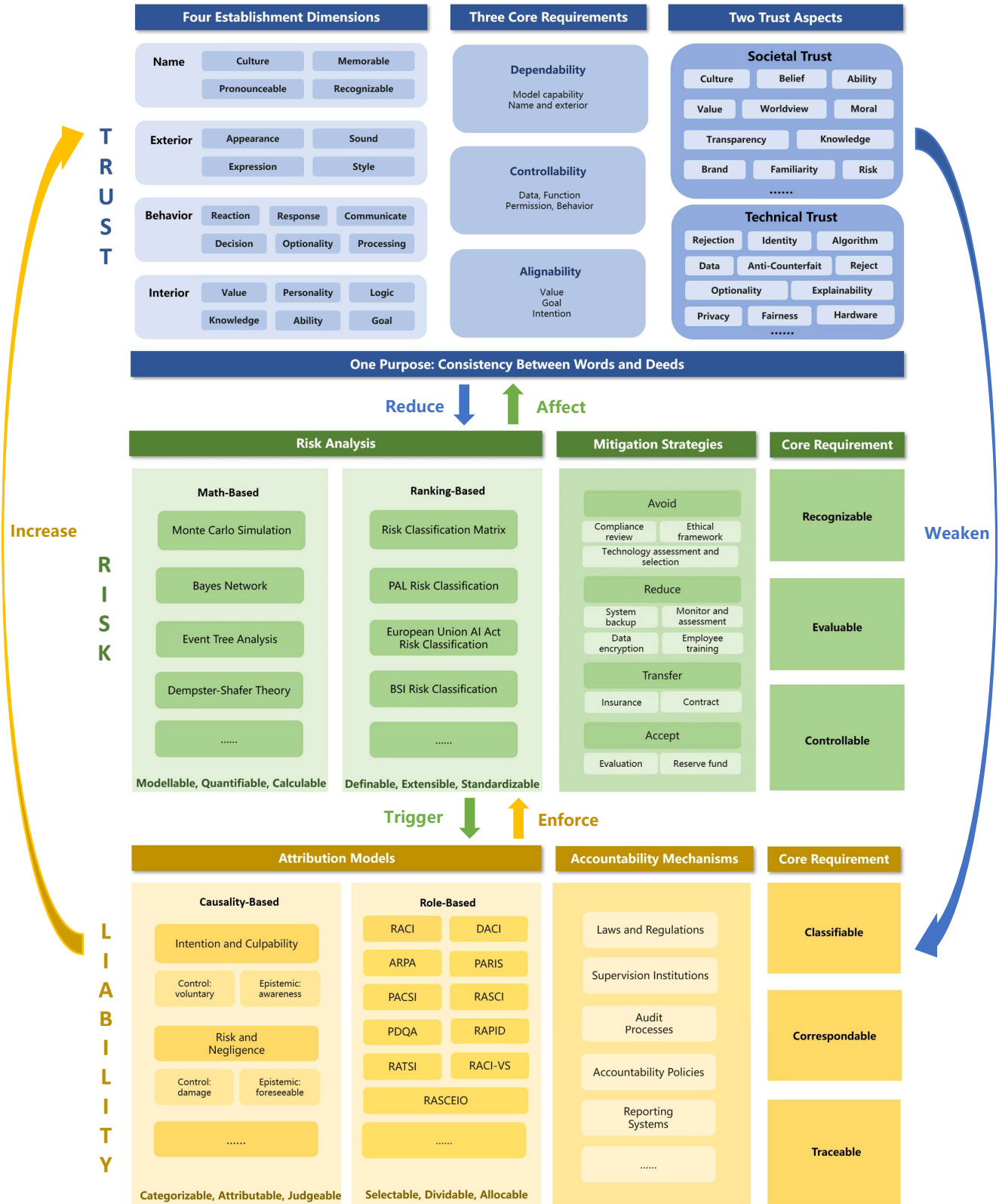

Figure 3. Trust system (blue), risk system (green), liability system (yellow).

## VI. LIABILITY SYSTEM

The liability system shown in Fig. 3 is an essential part of the TRL framework crucial for fostering trust and managing risks. The process involves allocating responsibilities to stakeholders using the attribution models and tracking the fulfillment of liabilities using the accountability mechanisms. Two types of attribution models are proposed in the liability system, with one based on role division, and the other based on causality.

### A. ROLE-BASED ATTRIBUTION MODELS

Most role-based attribution models evolve from the RACI model; the four letters in its acronym respectively refers to Responsible, Accountable, Consulted, and Informed. Stakeholders involved in the process will be assigned to one of the four roles, R, A, C, or I, based on their responsibility and contribution under different scenarios. Other models function similarly as RACI, with slight modifications in how roles are defined and arranged. After choosing a suitable attribution model, the next step is to match stakeholders to the corresponding roles with the help of a table. The columns are dedicated to different stakeholders involved, the rows show different scenarios, and the cells contain the role of each stakeholder under each scenario. Afterward, the third and last step is to allocate responsibilities to each stakeholder after considering their role and the risk level of the scenario.

### B. CAUSALITY-BASED ATTRIBUTION MODELS

On the other hand, the models based on causality involves a different attribution process. The first step is to choose an attribution model and identify related stakeholders. Options include intention-and-culpability-based models and risk-and-negligence-based models. Both models involve the definition of a control condition and an epistemic condition. In the intention-and-culpability-based model, the control condition refers to whether an agent "had a free and voluntary choice" concerning the action it took, while the epistemic condition requires the agent to have "actual knowledge and awareness of the particular facts" surrounding the action [16]. However, they are defined quite differently for a risk-and-negligence-based model. Here, the control condition showcases whether the agent "caused the relevant damage or injury," while the epistemic condition expects the agent to understand and avoid "reasonably foreseeable consequence" [16]. After choosing one model from the two options, the next step is to evaluate whether stakeholders in the scenarios meet the defined control condition and epistemic condition of the chosen model. Finally, responsibility will be allocated to each stakeholder based on the evaluation results determined in the previous step.

### C. ACCOUNTABILITY MECHANISMS

Whilst the attribution models assign responsibilities to stakeholders, the accountability mechanisms ensure that these liabilities are correctly bore and carried out by stakeholders. Accountability mechanisms listed in Fig. 3 differ in their magnitude of severity and level of authority, but they all serve the same purpose of ensuring issued liabilities are carried out correctly. Overall, the liability system is composed of the role-based and causality-based attribution models and the accountability mechanisms. Together, they ensure that liabilities are classifiable, correspondable, and traceable.

## VII. APPLYING THE TRUST, RISK, LIABILITY FRAMEWORK TO AI AGENTS

After introducing the structure and content of the trust system, risk system, and liability system in previous sections, this section provides examples and explanations that illustrate how to utilize the three systems and apply the TRL framework in the context of AI agents.

### A. UTILIZING THE TRUST SYSTEM

As described in the Trust System section, the trust system contains four establishment dimensions. These four trust dimensions specifically applied to the domain of AI agents will be unfolded in this section.

The first dimension is an AI agent's name, which can play a critical role in deciding users' first impressions. An easily memorizable, pronounceable, recognizable, and culturally significant name can help AI agents create decent first impressions and gain trust from users. The exterior of an AI agent plays an equally important role in influencing trust. One of the most important factors is facial expressions, including direct eye contacts, polite smiles, and appropriate nodding [17]. Sound also plays a role in influencing trust. Cornell University's study shows that varying volumes, tones, and occasional emphasis will increase listener's perceived authenticity of information and hence their trust [18]. The interior dimension will start to affect trust as the users began to interact with the AI agent. Quality of data, capability of the model, transparency and explainability, rejection models, value system, possession of human traits, and privacy protection are all determining factors [19] [20]. An interior dimension characterized by professionality, accuracy, explainability, with personified characteristics, strong model capabilities, clear privacy protection measures, and alignable values will be important for increasing the trustworthiness of an AI agent. The last dimension that can influence trust is the AI agent's behavior. The comprehensibility of decisions and their alignability with users' expectations, the timeliness, convenience, and accuracy of responses, as well as the communication methods developed and tailored to specific users will all enhance users' experiences of interacting with the AI agents and hence the trust towards the AI agents. A transparent and controllable AI agent, with iterative update and alignment with users' expectations, constitute an ideal behavior dimension.

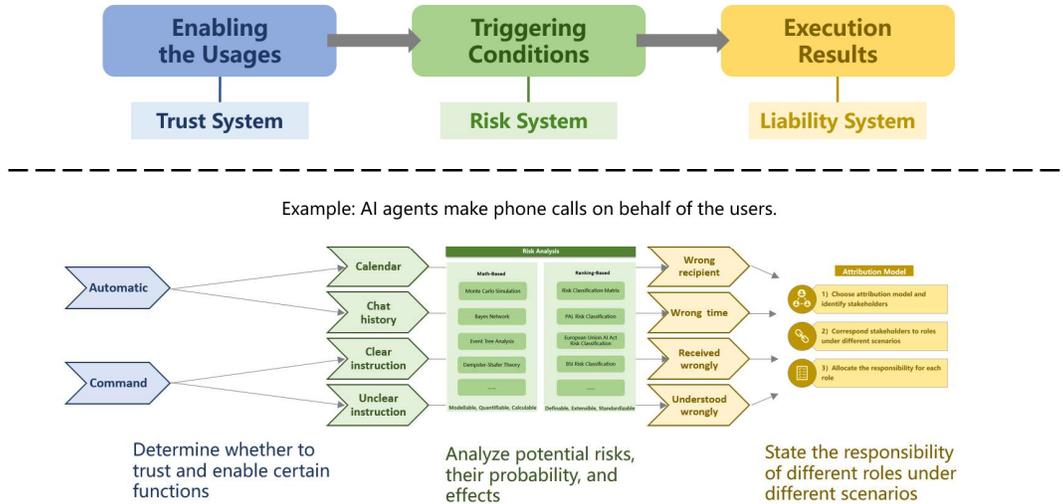

Figure 4. Applying the Trust, Risk, Liability Framework.

These four dimensions of the trust system determine users' degrees of trust towards AI agents, which in turn decide whether they will activate and use certain functionalities of the AI agents. Afterward, it is up to the risk system and liability system to determine foreseeable risks of the functionalities and allocate liabilities for potential outcomes and consequences, as shown in Fig. 4.

### B. UTILIZING THE RISK SYSTEM

A hypothetical scenario, where the AI agent picks up and makes phone calls on the user's behalf, will be used to explain how to utilize the risk system in the context of AI agents. The action of calling someone may be triggered through several different ways. Each scenario involves different risks, probabilities, and potential consequences.

The first step of analyzing the risks involved is to choose a risk ranking model and set risk indicators accordingly. Privacy, confidentiality of data, and auditability are examples of possible indicators. A matrix will be used to rank risks associated with different events. The columns are the indicators and the rows are the events. Depending on whether each indicator is satisfied, a checkmark will be inserted into the corresponding cells, and an overall risk level will be determined for each row.

Level 0 corresponds to the lowest risk, where risk acceptance may be used and AI agents should be allowed to make phone calls independently. Level 1 needs risk reduction, so the user's past call histories should be analyzed to derive safer options. Risk transfer will be used for level 2, such that AI services with sufficient insurance policies should be chosen. The highest risk corresponds to level 3, and the most appropriate action is to adopt risk avoidance and give up the calling plan.

### C. UTILIZING THE LIABILITY SYSTEM

Within the same phone call example, the liability system is applied for the execution stage of the functionalities, to specify the responsibilities and liabilities of each stakeholder under different possible results. Assume the user sends a command for the AI agent to call a designated person, but the command was unfortunately wrongly received by the AI agent.

The first step outlined in the liability system is to choose an attribution model, like the DACI model. A table should be constructed to correspond stakeholders with their respective roles. The four columns are AI agent, user, company, and call recipient, and one of the rows correspond to the phone call command. Driver, Approver, Contributor, and Informed will be respectively filled in for these four stakeholders. As the Driver of the action, the AI agent is expected to take the largest share of responsibilities. The user, as the approver of the action, did not recognize mistakes in time and clarify accordingly, and thus also need to bear consequences. The company manufacturing the AI agent will be the contributor. If the wrongly received command originated from apparent loopholes in the algorithms, and if such codes resulted in severe consequences, the company may be liable as well. Lastly, recipients of the phone calls are the informed ones. If they discovered obvious abnormalities in the conversation, they should verify with the users to prevent discussed actions from taking place.

Overall, the trust, risk, and liability systems within the TRL framework are utilized successively to determine whether to trust and enable functionalities, to analyze risks and consequences, and to state responsibilities of stakeholders for each application scenario.

## VIII. IMPACTS OF THE TRUST, RISK, LIABILITY FRAMEWORK

The values and implications of the TRL framework lie in its potential societal impacts, economic impacts, ethical impacts, and legal impacts.

Firstly, the TRL framework can bring remarkable impacts to society. By following the trust principles and requirements outlined by the trust system, AI developers can increase the users' overall trust level and eventually build public trust in AI. Meanwhile, the risk system can assist the developers to address high-risk components and guide the users to avoid taking high-risk actions. The accountability mechanisms can ensure that wrongful actions will receive corresponding consequences, promoting AI usage to users and industries who were previously concerned about the lack of liability attribution. Together, the TRL framework effectively contributes to reducing the fear or resistance towards using AI agents that originated from privacy and trustworthiness concerns. With a higher level of recognition and higher public trust, AI can be integrated into all aspects of society more smoothly and effectively.

At the same time, the TRL framework also carries great economic impacts. The accountability mechanisms in the liability system can help to ensure that AI agents accept and bear relevant consequences; more liable AI agents reduces legal disputes and the associated costs. Moreover, the trust system and the liability system contribute to fostering a safer, more reliable, and more trustworthy AI agent. These characteristics will make businesses more willing to incorporate AI agents into their operations and make consumers more willing to use the products, opening up the market for AI and enlarging the economic gains.

Moreover, the TRL framework is also capable of bringing ethical and political impacts. With a better sense of responsibility, AI agents can decrease the number of incidents and harms towards the users and protect the users' rights and interests. As a solution that paves the way for addressing issues related to trust, risks, and liability, it has the potential to shape future regulations and lead to the establishment of more robust AI policies.

Overall, the TRL framework has ample opportunities and prospects to leave significant impacts in the near future.

## IX. CONCLUSION

This paper proposes and introduces the TRL framework as a solution to address the trust issues, risks, and liability attribution struggles related to the usage of AIs. The TRL framework consists of a trust, risk, and liability system with interdependent influences. The liability system tackles responsibility allocation methods and accountability mechanisms; the risk system includes risk analysis and corresponding mitigation strategies; and the trust system details the establishment dimensions, requirements, aspects, and the ultimate principle of trust. Together, the TRL framework may be applied to any specific application scenarios to analyze and take appropriate measures with regard to the context. It is expected to leave considerable impacts in societal, economic, ethical, and legal aspects. However, as a newly proposed framework, efforts still need to be taken to ensure a successful and effective launch and promotion. After clearing potential obstacles along the way, the TRL framework has the potential to contribute remarkably to developing and utilizing trustworthy, risk-free, and responsible AIs in 6G networks.

## BIOGRAPHIES

**Jiayun Mo** (jmo005@e.ntu.edu.sg) is currently pursuing the Bachelor of Science degree in Computer Science at Nanyang Technological University, Singapore (NTU). During her internship at Shield Laboratory, the Singapore Research Center of Huawei International Pte. Ltd., she delved into research about trust modeling and its related applications. She intends to aim for further studies in order to better pursue her current research interests of human-centered applications of technologies.

**Xin Kang** (kang.xin@huawei.com) is senior researcher at Huawei Singapore Research Center. Dr. Kang received his Ph.D. Degree from National University of Singapore. He has more than 15 years' research experience in wireless communication and network security. He is the key contributor to Huawei's white paper series on 5G security. He has published 70+ IEEE top journal and conference papers, and received the Best Paper Award from IEEE ICC 2017, and Best 50 Papers Award from IEEE GlobeCom 2014. He has also filed 60+ patents on security protocol designs, and contributed 30+ technical proposals to 3GPP SA3. He is also the initiator and chief editor for ITU-T standard X.1365, X.1353, Y.3260, and the on-going work item Y.Trust-AI.

**Tieyan Li** (li.tieyan@huawei.com) is currently leading Digital Trust research, on building the trust infrastructure for future digital world, and previously on mobile security, IoT security, and AI security at Shield Lab., Singapore Research Center, Huawei Technologies. Dr. Li is also the director of Trustworthy AI C-TMG and the vice-chairman of ETSI ISG SAI. Dr. Li received his Ph.D. Degree in Computer Science from National University of Singapore. He has more than 20 years experiences and is proficient in security design, architect, innovation and practical development. He was also active in academic security fields with tens of publications and patents. Dr. Li has served as the PC members for many security conferences, and is an influential speaker in industrial security forums.

**Zhongding Lei** (lei.zhongding@huawei.com) is currently a Senior Researcher with the Huawei Singapore Research Center. He has been working on 5G network security, since 2016. Prior to joining Huawei, he was the Laboratory Head and a Senior Scientist with the Agency for Science, Technology, and Research (A-STAR), Singapore, involved in research and development of 3GPP and IEEE standards in wireless systems and networks. He has been the Editor-in-Chief of IEEE Communications Standards Magazine, since 2019.